\documentstyle[amstex,twoside,fleqn,npb
]{article}
\title{Construction of nonlocal light--cone operators
with definite twist\thanks{Contribution to 7th Intern. Workshop
on Deep Inelastic Scattering and QCD, Zeuthen, April 1999;
to apppear: Nucl. Phys. B (Proc. Suppl.)
}}

\author{B. Geyer$^a$, M. Lazar\address{
     Institut f\"ur Theoretische Physik, Universit{\"a}t Leipzig,
        Augustusplatz 10, D--04109 Leipzig\\
$^b$~DESY-IfH Zeuthen, Platanenallee 6, D--15735 Zeuthen} %
        and
        D. Robaschik$^b$}

\begin{document}


\maketitle

\def \slas{\kern -7.2pt /}
\def \sla{\kern -5.4pt /}
\def \Cslas{\kern -6.8pt /}
\def \Dslas{\kern -7.4pt /}
\def \ii{\mathrm{i}}
\def \d{\mbox{d}}
\def \dd#1{\frac{\mathrm{d}}{\mathrm{d}#1}}
\def \lcd{\tilde{\partial}}
\def \pd{\partial}
\def \e{\mbox{e}}
\def \lcxg{x\slas}
\def \lcx{\tilde{x}}
\def \lcxs{\tilde{x}\slas}
\def \tl#1{\overset{\kern 0pt\circ}{#1}}
\def \TL#1{\overset{\kern -3pt \circ}{#1}}
\def \relstack#1#2{\mathrel{\mathop{#2}\limits_{#1}}}
%
\newcommand{\equ}[2]{\begin{equation}\label{e#1}#2\end{equation}}
\newcommand{\equa}[2]{\begin{eqnarray}\label{e#1}#2\end{eqnarray}}
\newcommand{\abl}[2]{\frac{\delta #1}{\delta#2}}
\newcommand{\D}{\displaystyle}
\newcommand{\mr}{\mathrm}
\newcommand{\gam}{\mathrm{\Gamma}}
\newcommand{\tb}[1]{\textbf{#1}}
\renewcommand{\(}{\left(}
\renewcommand{\)}{\right)}
\renewcommand{\[}{\left[}
\renewcommand{\]}{\right]}
\renewcommand{\~}{\tilde}
\renewcommand{\^}{\hat}
\renewcommand{\phi}{\varphi}
\renewcommand{\theta}{\vartheta}
\newcommand{\lamb}{\bar\lambda_k}
\newcommand{\lam}{\lambda_k}
\newcommand{\del}{\bigtriangleup}
\newcommand{\Z}{Z_{Nk}}
\newcommand{\Gb}{\bar{G}}
\newcommand{\bq}{\begin{equation}}
\newcommand{\eq}{\end{equation}}
\newcommand{\bea}{\begin{eqnarray}}
\newcommand{\eea}{\end{eqnarray}}
\newcommand{\gap}{\stackrel{>}{\sim}}
\newcommand{\lap}{\stackrel{<}{\sim}}
\newcommand{\fem}{$f_2^{em}$}
\newcommand{\qsq}{$Q^2$}
\newcommand{\alsq}{$\alpha_s(Q^2)$}
\newcommand{\dals}{$\delta\alpha_s$}
\newcommand{\pbinv}{pb^{-1}}
\newcommand{\lsim}{\raisebox{-0.07cm}{$\,\stackrel{<}{{\scriptstyle\sim}}\,$}}
\newcommand{\gsim}{\raisebox{-0.07cm}{$\,\stackrel{>}{{\scriptstyle\sim}}\,$}}
\newcommand{\dfeq}{\stackrel{\scriptsize =}{\scriptsize df}}
\newcommand{\als}{\alpha_s}
\newcommand{\aqu}{\langle Q^2\rangle}
\newcommand\pB{[\mbox{pb}]}
\newcommand\MSbar{$\overline{\mbox{MS}}$}
\newcommand\order{{\cal O}}
\newcommand\mat{{\cal M}}
\newcommand\ds{\displaystyle}
\newcommand\MeV{\,\mbox{MeV}}
\newcommand\GeV{\,\mbox{GeV}}
\newcommand\TeV{\,\mbox{TeV}}
\newcommand\secu{\,\mbox{sec}}
\newcommand\kvec{\mbox{\boldmath $k$}}
\newcommand\MT{Morfin and Tung}
\newcommand\CALL{{\cal L}}
\newcommand\KG{\kappa_G}
\newcommand\KA{\kappa_A}
\newcommand\LG{\lambda_G}
\newcommand\LA{\lambda_A}
\newcommand\PD{\not p}
\newcommand\MP{M^2_{\Phi}}
\newcommand\B{\beta}
\newcommand\XL{\log \left|\frac{1 + \B}{1 -\B}\right |}
\newcommand\SM{\frac{\hat{s}}{M_{\Phi}^2}}
\newcommand\SMM{\frac{\hat{s}^2}{M_{\Phi}^4}}
\newcommand\SMMM{\frac{\hat{s}^3}{M_{\Phi}^6}}
\newcommand\SMMMM{\frac{\hat{s}^4}{M_{\Phi}^8}}
\newcommand\BC{\beta^2 \cos^2 \theta}
\newcommand\CB{\beta^2 \cos^2 \theta}
\newcommand\CBB{\beta^4 \cos^4 \theta}
\newcommand\CBBB{\beta^6 \cos^6 \theta}
\newcommand\CBBBB{\beta^8 \cos^8 \theta}
\newcommand\SH{\hat{s}}
\newcommand\uh{\hat{u}}
\newcommand\thh{\hat{t}}
\newcommand\sh{\hat{s}}
\newcommand\ea{\varepsilon_1}
\newcommand\eb{\varepsilon_2}
\newcommand\Ea{\epsilon_1}
\newcommand\Eb{\epsilon_2}
\newcommand\ch{{\rm{ch}}}
\newcommand\shh{{\rm{sh}}}
\newcommand\thhh{{\rm{th}}}
\newcommand\st{\sin^2 \theta}
\newcommand\BR{\left(1 - \CB\right)}
\newcommand\ka{\kappa_1}
\newcommand\kb{\kappa_2}
\newcommand\kap{\kappa_1'}
\newcommand\kbp{\kappa_2'}
\newcommand\xx{\tilde{x}}
\newcommand\AAA{\alpha_1}
\newcommand\AB{\alpha_2}
\newcommand\Kvec{\mbox{\boldmath$K$}}
\newcommand\Pvec{\mbox{\boldmath$P$}}
\newcommand\Uvec{\mbox{\boldmath$U$}}
\newcommand\Vvec{\mbox{\boldmath$V$}}
\newcommand\bx{\overline{x}}
\newcommand\by{\overline{y}}
\newcommand\FFA{\mbox{$\widetilde{F}^a$}}

\section{Introduction}
With the growing precision of data for light--cone dominated
hard scattering processes like DIS, DVCS and various (semi) exclusive
processes a better perturbative understanding of QCD concerning
higher twist contributions is required.
Thereby the {\em nonlocal} light--cone expansion (LCE) is
optimally adapted since the {\em same} nonlocal LC operator,
and its anomalous dimension, are related to {\em different}
phenomenological distribution amplitudes, and their
$Q^2$--evolution kernels \cite{AZ,MRGDH}. But,
`geometric' twist  $=$ dimension $-$
spin, $\tau=d-j$, introduced 
for the {\em local} LC--operators \cite{GT} cannot be
extended directly to the nonlocal LC--operators.
On the other hand,
motivated by LC--quantization where the quark fields may be
decomposed into `good' and `bad' components \cite{KS} 
and by kinematic phenomenology \cite{JJ} 
the notion of `dynamic' twist $(t)$ was introduced counting
powers $Q^{2-t}$ of the momentum transfer. However, this
notion is defined only for {\em matrix elements} of operators, is
{\em not} Lorentz invariant and its relation to `geometric' twist
is complicated,~cf.~\cite{BBKT}.

Here, we introduce a systematic procedure to uniquely decompose
nonlocal LC--operators into harmonic operators of well defined
geometric twist, cf.~Ref.~\cite{GLR}. This will be demonstrated for the
case of (pseudo)scalar, (axial) vector and (skew) tensor
bilocal quark light--ray operators.

\section{General procedure}

Let us shortly state the consecutive steps which are
used to decompose a bilocal light--ray operator into operators
of definite twist:\\
(1) {\em Expansion} of the nonlocal operators for {\em arbitrary}
values of $x$ 
into a Taylor series of {\em local} tensor operators
(rank $n$, dimension $d$); cf.~Eq.~(\ref{GLCEOP}).\\
(2) {\em Decomposition} of these local operators
w.r.t {\em irreducible
tensor representations} of the (ortho\-chronous)
Lorentz group having definite rank, dimension
and spin $(n, d, j)$.\\
(3) {\em Resummation} of the infinite series (for any $n$)
of irreducible tensor operators of equal twist $\tau$
to nonlocal {\em  harmonic operators of definite twist}.\\
(4) {\em Projection} onto the light--cone,
$x\rightarrow\xx$, with
\begin{eqnarray}
\label{LCV}
\xx = x + \eta
{(x\eta)}
\big(
\sqrt{1 - {x^2}/{(x\eta)^2}} - 1
\big),~~ \eta^2=1,
\nonumber
\end{eqnarray}
leading to the required
twist decomposition:
\begin{eqnarray}
\label{LCdecomp}
O_\Gamma(\kappa_1 \tilde x,\kappa_2\tilde{x})
~=&
\sum_{\tau_i = \tau_{\rm min}}^{\tau_{\rm max}}
O_\Gamma^{\tau_i}(\kappa_1 \tilde x,\kappa_2\tilde{x}).
\end{eqnarray}

Generically the nonlocal quark operators
with tensor structure
$
\Gamma=\{1,\gamma_\alpha,\sigma_{\alpha\beta},
\gamma_5\gamma_\alpha,\gamma_5\}
\nonumber
$
for arbitrary $x$ are represented by
\begin{eqnarray}
\label{GLCEOP}
\lefteqn{
O_\Gamma(\ka x, \kb x)
=
\;:\overline\psi(\ka x) \Gamma
U(\ka x, \kb x)
\psi(\kb x):}
\nonumber\\
&=
\sum_{n=0}^{\infty}
(n!)^{-1}
{x}^{\mu_1}\ldots
{x}^{\mu_n}
O^{\Gamma}_{\mu_1\ldots\mu_n}(\ka,\kb),
\end{eqnarray}
with the corresponding local operators
\begin{eqnarray}
\label{local}
O^{\Gamma}_{\mu_1\ldots\mu_n}
\equiv
\left[\bar{\psi}(x)\Gamma
{\sf D}_{(
\mu_1}\!\ldots{\sf D}_{\mu_n)}\!
\psi (x)\right]_{x=0},
\end{eqnarray}
where the brackets $(\ldots)$ denote total symmetrization;
thereby we used the phase factors
\begin{eqnarray}
\label{phase}
U(\ka x, \kb x)
\!\!\!&=& \!\!\!
{\cal P}
\exp\Big\{-ig
\int^{\ka}_{\kb}\!\! d\kappa' \,x^\mu A_\mu (\kappa' x)
\Big\},
\nonumber
\end{eqnarray}
and the generalized covariant derivatives
\begin{eqnarray}
\label{D_kappa}
{\sf D}_\mu(\kappa_1,
\kappa_2)\!\!\!
&\equiv& \!\!\!
\kappa_1 ({\stackrel{\leftarrow}{\pd}}_\mu-i g A_\mu)+
\kappa_2 ({\stackrel{\rightarrow}{\pd}}_\mu+i g A_\mu).
\nonumber
\end{eqnarray}

Concerning the decomposition of the local tensor operators
(\ref{local}) into irreducible tensors w.r.t.~the
 Lorentz group $\cal{L}^\uparrow$ we remember that these are
uniquely given by {\em traceless} tensors being classified
according to their {\em symmetry class} w.r.t.~the symmetric
group. A symmetry class is determined by a (normalized)
{\em  Young-Operator}
 ${\cal Y}_{[m]}={f_{[m]}}{\cal PQ}/{n!}$,
where $[m]= (m_1\geq m_2\geq\ldots\geq m_r)$ with
$\sum^r_{i=1}m_i = n$ defines a Young pattern,
 ${\cal P}=\sum_{p\in H_{[m]}}p$ and
${\cal Q}=\sum_{q\in V_{[m]}}\delta_q\, q$ are the
symmetrizations and antisymmetrizations, respectively,
related to the horizontal ($H_{[m]}$) and vertical ($V_{[m]}$)
permutation w.r.t.~a standard tableau
obtained by putting into $[m]$ the numbers $1, \ldots, n$
raising from left to right and from top to bottom;
$f_{[m]}$ is the number of standard tableau's to $[m]$.

For the Lorentz group (in $4$ dimensions)
the allowed Young patterns are restricted by
$\ell_1+\ell_2 \leq 4~ (\ell_i:$ lenght of columns).
Since the local operators (\ref{local}) are
totally symmetric w.r.t.
$\mu_i$'s, depending to the additional tensor
structure $\Gamma$, only the following Young patterns
are of relevance (the spin $j$ is
related to the various trace terms):\\
$
\begin{array}{lll}
\hspace{-.2cm}
\rm{(i)}&[m] = (n)&j = n, n-2, n-4,...\\
\hspace{-.2cm}
\rm{(ii)}&[m] = (n-1,1)&j = n-1, n-2,...\\
\hspace{-.2cm}
\rm{(iii)}&[m] = (n-2,1,1)&j = n-2, n-3,...\\
\end{array}
$
\vspace*{0cm}
\hspace*{.3cm} Because the decomposition into irreducible
representations does not depend on the specific values of
the $\kappa$--variables we may choose
$(\kappa_1,\kappa_2)=(0,\kappa)$ and
${\sf D}_\mu(\kappa_1, \kappa_2)
\Rightarrow \kappa {\stackrel{\rightarrow}{D}}_\mu (A)$;
afterwards we generalize to arbitrary values of $\kappa_i$.

\section{Harmonic tensor functions}
For constructing
traceless tensors $\tl T_{\Gamma (\mu_1 \ldots \mu_n)}$
let us start with a generic tensor, not being traceless, whose
symmetrized indices are contracted by $x^{\mu_i}$'s:
 $T_{\Gamma n}(x) =
x^{\mu_1}\ldots x^{\mu_n} T_{\Gamma(\mu_1 \ldots \mu_n)}$.
The conditions for these tensors to be traceless read:
\begin{eqnarray}
\square \tl T_{\Gamma~n}(x)=0,
\end{eqnarray}
and in addition
\begin{eqnarray}
\pd^\alpha \tl T_{\alpha~n}(x) = 0
 \quad{\rm resp.}\quad
\pd^\alpha \tl T_{[\alpha\beta]~n}(x) = 0
\end{eqnarray}
for (axial) vectors resp. skew
tensors.\\
The solutions of these
equations ($D=4$) are:\\
(1) {\em Scalar harmonic polynomials}:
\begin{eqnarray}
\label{T_harm4}
\tl T_n(x)
=
H^{(4)}_n\!\left(x^2|\square\right)T_n(x)
\end{eqnarray}
with  the harmonic projection operator 
\begin{eqnarray}
\label{Harm4}
H^{(4)}_n\!\left(x^2|\square\right)
=
\sum_{k=0}^{[\frac{n}{2}]}
\frac{(n-k)!}{k!n!}
\left(\frac{-x^2}{4}\right)^{\!\!k}
\square^{k}.
\end{eqnarray}
(2) {{\em Vectorial harmonic polynomials}:
\begin{eqnarray}
\label{T_vecharm}
\tl T_{\alpha n}(x)
\!\!\!\!&=&\!\!\!\!
\Big\{\delta_{\alpha}^{\beta}
-\hbox{\Large$\frac{2}{(n+1)^2}$}
\Big(x_\alpha\pd^\beta(x\pd)
\nonumber\\
\!\!\!\!&&\!\!\!\!
-\hbox{\Large$\frac{x^2}{2}$}
\pd_\alpha\pd^\beta\Big)\!\Big\}
H^{(4)}_n\!\left(x^2|\square\right) T_{\beta n}(x).
\end{eqnarray}
Of course, contraction with $x^\alpha$ leads to $\tl T_n(x)$.\\
(3) {{\em Skew tensorial harmonic polynomials}:
\begin{eqnarray}
\label{T_tenharm}
\lefteqn{
\tl T_{[\alpha\beta]n}(x)
=
\Big\{\delta_{[\alpha}^\mu\delta_{\beta]}^\nu
+\hbox{\Large$\frac{1}{(n+1)n}$}
\Big(
x_{[\alpha}\delta_{\beta]}^{\mu}\pd^{\nu}(x\pd)}
\nonumber\\
&-\hbox{\Large$\frac{x^2}{2}$}
\pd
_{[\alpha}\delta_{\beta]}^{\mu}\pd^{\nu}\Big)
-\hbox{\Large$\frac{2}{(n+2)(n+1)n}$}
x_{[\alpha}\pd_{\beta]}x^{\mu}\pd^{\nu}
\Big\}
\nonumber\\
&
\times
H^{(4)}_n\!\left(x^2|\square\right)
T_{[\mu\nu] n}(x)
\end{eqnarray}
with the convention
$T_{[\alpha \beta]} = (T_{\alpha\beta} - T_{\beta\alpha})/2$.\\
In principle also harmonic tensors like $\tl T_{(\alpha\beta)n}(x)$,
related to gluonic operators, and of higher orders may be constructed.
Obviously, any {\em harmonic tensor function} may be represented
as a series expansion into corresponding polynomials.

\section{Twist decomposition of quark operators}

In order to select the different spin content of these
tensorial harmonic polynomials we have to observe the complete
symmetry  related to $[m]$. Let us give some selected examples.

\noindent{\em 1. (Pseudo) scalar operators}:
First we consider the scalar quark operator
\begin{eqnarray}
\label{O2}
\hspace{-.7cm}
&&O(0,\kappa x)=
\bar{\psi}(0) (\gamma x)
U(0,\kappa x)\psi(\kappa x)
\\
\hspace{-.7cm}
&&O_n(x)=
x^\alpha x^{\mu_1}\ldots x^{\mu_n}
\bar{\psi}(0)
\gamma_{(\alpha}
D_{\mu_1}\ldots D_{\mu_n)}\psi(0).
\nonumber
\end{eqnarray}
Only Young pattern (i) is relevant:\\
\\
\unitlength0.4cm
\begin{picture}(20,1)
\linethickness{0.15mm}
\put(1,0){\framebox(1,1){$\mu_1$}}
\put(2,0){\framebox(1,1){$\mu_2$}}
\put(3,0){\framebox(3,1){$\ldots$}}
\put(6,0){\framebox(1,1){$\mu_n$}}
\put(8,0){contains}
\put(13,0){${\tau=2},{4},\ldots~.$}
\\
\end{picture}
\vspace*{.0cm}

\noindent
Therefore, $O_n(x)$ gets traceless by applying,
according to Eq.~(\ref{T_harm4}), the harmonic
projection operator. Using the integral representation
of Euler's Beta--function, $(n+1-k)!(k-1)!/(n+1)!
=B(n+2-k,k)=\int_0^1\d t\, t^{n+1-k}(1-t)^{k-1}$ one


\noindent resums
these local operators $\tl O_{n}(x)$ to obtain
the traceless nonlocal scalar twist 2 operator
\begin{eqnarray}
\label{proj_tw2}
\lefteqn{
O^{\rm tw2}(0,\kappa x)\equiv
\tl O(0,\kappa x)=
\sum_{n=0}^\infty \frac{\kappa^n}{n!}\tl O_n(x)
= }
\\
\!\!\!\!\!\!\!\!&&\!\!\!\!\!\!\!\!
O(0,\!\kappa x)\!+\!
\sum_{k=1}^{\infty}\!\int_0^1\!\!\!\!d t
\frac{(1\!-\!t)^{k-1}(-x^2)^k\square^k}{t^{k-1}4^k k!(k-1)!}
O(0,\!\kappa tx).
\nonumber
\end{eqnarray}
The sum contains the trace terms
which are to be subtracted from $O(0,\kappa x)$;
obviously, they vanish in the limit
$x \rightarrow \xx$. Therefore, the scalar
operator (\ref{O2}), if taken on LC, already has minimal twist 2.
The analogous conclusion holds for the pseudo scalar
operator obtained by replacing
$\gamma_\mu \rightarrow \gamma_5\gamma_\mu$.

\noindent{\em 2. (Axial) vector operators}:
Now we consider the vector operators
\begin{eqnarray}
\label{O_ent}
O_{\alpha}(0,\kappa\tilde{x})
\!\!\!\!&=&\!\!\!\!
\bar{\psi}(0)
\gamma_{\alpha}
U(0,\kappa\tilde{x})\psi(\kappa\tilde{x})
\\
O_{\alpha n}(x)
\!\!\!\!&=&\!\!\!\!
x^{\mu_1}\ldots x^{\mu_n}
\bar{\psi}(0) \gamma_\alpha
D_{(\mu_1}\ldots
D_{\mu_n)} \psi (0)
\nonumber
\end{eqnarray}
The relevant symmetry classes are determined by the
Young patterns (i) and (ii):
\\
\\
\unitlength0.4cm
\begin{picture}(20,1)
\linethickness{0.15mm}
\put(1,0){\framebox(1,1){$\alpha$}}
\put(2,0){\framebox(1,1){$\mu_1$}}
\put(3,0){\framebox(1,1){$\mu_2$}}
\put(4,0){\framebox(3,1){$\ldots$}}
\put(7,0){\framebox(1,1){$\mu_n$}}
\put(9,0){contains}
\put(13,0){${\tau=2},{4},\ldots,$}
\end{picture}
\\
\vspace*{1mm}
\unitlength0.4cm
\begin{picture}(20,1)
\linethickness{0.15mm}
\put(1,-1){\framebox(1,1){$\alpha$}}
\put(1,0){\framebox(1,1){$\mu_1$}}
\put(2,0){\framebox(1,1){$\mu_2$}}
\put(3,0){\framebox(3,1){$\ldots$}}
\put(6,0){\framebox(1,1){$\mu_n$}}
\put(9,0){contains}
\put(13,0){${\tau=3},{4},\ldots,$}
\\
\end{picture}
\vspace*{2mm}

\noindent
Proceeding analogous to the scalar case
we obtain for the symmetry class (i)
\begin{eqnarray}
\label{O2i}
\hspace{-.7cm}
&&O^{\mathrm{tw2}}_{\alpha}(0,\kappa{x})
=\pd_\alpha
\int_{0}^{1} \!\! \d\lambda\;
\TL{O}(0,\kappa\lambda{x}),
\\
\hspace{-.7cm}
&&{\rm and~ for~ the~ symmetry~ class~ (ii)}
\\
\label{O3ii}
\hspace{-.7cm}
&&O^{\mathrm{tw3}}_{\alpha} (0,\kappa x)
=
\left(g_{\alpha}^{\mu}(x\pd)-x^\mu\pd_\alpha\right)
\!\!\int_{0}^{1}\!\!
\d\lambda
\TL{O_\mu}(0,\kappa\lambda x),
\vspace{-.3cm}
\nonumber
\end{eqnarray}
where $\TL{O_\alpha}(0,\kappa x)$ has to be determined
according to Eq.~(\ref{T_vecharm}).
Both expressions contain trace terms which are related
to twist $\tau = 4$ and higher.

\noindent{\em 3. (Skew) tensor operators}:
Here we distinguish two cases, the true tensor
case (A)
\begin{eqnarray}
\hspace{-.7cm}
\label{O_ent}
&&M_{[\alpha\beta]}(0,\kappa{x})
=
\bar{\psi}(0)
\sigma_{\alpha\beta}
U(0,\kappa{x})\psi(\kappa{x}),
\\
\hspace{-.7cm}
&&M_{[\alpha\beta]n} (x)
=
x^{\mu_1}...~x^{\mu_n}
\bar{\psi}(0) \sigma_{\alpha\beta}
D_{(\mu_1}...~D_{\mu_n)} \psi(0),
\nonumber
\end{eqnarray}
and case (B) with one additional contraction
\begin{eqnarray}
M_{\alpha}(0,\kappa{x})
\!\!\!\!&=&\!\!\!\!
\bar{\psi}(0)
\sigma_{\alpha\beta}\xx^\beta
U(0,\kappa{x})\psi(\kappa{x}),
\\
M_{\alpha n} (x)
\!\!\!\!&=&\!\!\!\!
x^\beta M_{[\alpha\beta]n} (x).
\nonumber
\end{eqnarray}
The relevant symmetry classes are determined
by the Young patterns (ii) and (iii):
\\
\\
\unitlength0.4cm
\begin{picture}(20,1)
\linethickness{0.15mm}
\put(1,-1){\framebox(1,1){$\alpha$}}
\put(1,0){\framebox(1,1){$\beta$}}
\put(2,0){\framebox(1,1){$\mu_1$}}
\put(3,0){\framebox(1,1){$\mu_2$}}
\put(4,0){\framebox(3,1){$\ldots$}}
\put(7,0){\framebox(1,1){$\mu_n$}}
\put(9,0){contains}
\put(13,0){$\tau=2, 3,4\ldots,$}
\end{picture}
\\
\vspace*{4mm}
\unitlength0.4cm
\begin{picture}(20,1)
\linethickness{0.15mm}
\put(1,-2){\framebox(1,1){$\alpha$}}
\put(1,-1){\framebox(1,1){$\beta$}}
\put(1,0){\framebox(1,1){$\mu_1$}}
\put(2,0){\framebox(1,1){$\mu_2$}}
\put(3,0){\framebox(3,1){$\ldots$}}
\put(6,0){\framebox(1,1){$\mu_n$}}
\put(9,0){contains}
\put(13,0){$\tau=3,4,\ldots~.$}
\end{picture}
\vspace*{5mm}

\noindent
Obviously, in case (B) the Young pattern (iii) does
not contribute.
From pattern (ii) we obtain
\begin{eqnarray}
\label{M2ii}
M^{\mathrm{tw2}}_{[\alpha\beta]}(0,\kappa{x})
\!\!\!\!&=&\!\!\!\!
2\pd_{[\beta}\int_{0}^{1} \d\lambda\,\lambda\,
\TL{M_{\alpha]}}(0,\kappa\lambda{x}),
\\
M^{\mathrm{tw2}}_{\alpha}(0,\kappa{x})
\!\!\!\!&=&\!\!\!\!
\TL{M_{\alpha}}(0,\kappa\lambda{x});
\end{eqnarray}
the contributions from pattern (iii) begin with
\begin{eqnarray}
\label{M3iii}
M^{\mathrm{tw3(iii)}}_{[\alpha\beta]}(0,\kappa{x})
\!\!\!\!&=&\!\!\!\!
\Big((x\pd)\delta^\nu_{[\beta}
- 2 x^\nu \pd_{[\beta}\Big)\times
\nonumber\\
&&
\int_{0}^{1} \d\lambda\,\lambda\,
\TL{M}_{\alpha]\nu}(0,\kappa\lambda{x}),
\vspace*{-.3cm}
\end{eqnarray}
\vspace{-.0cm}
where $\TL{M_{\alpha\nu}}(0,\kappa{x})$
is determined by Eq.~(\ref{T_tenharm}).

\begin{table*}[hbt]
\hrulefill
\begin{eqnarray}
\label{O2k}
O^{\rm tw2}(\ka\xx,\kb\xx)
\!\!&=&\!\!
\bar{\psi}(\ka\lambda\xx)(\gamma\xx)
U(\ka\lambda\xx, \kb\lambda\xx) \psi(\kb\lambda\xx)
\equiv
\xx^\mu O_\mu(\ka\xx,\kb\xx)
\end{eqnarray}
\begin{eqnarray}
\label{O2vec}
O^{\mathrm{tw2}}_{\alpha}(\ka\xx,\kb\xx)
\!\!&=&\!\!
\int_{0}^{1} \!\!\d\lambda
\Big(\pd_\alpha +
\hbox{\Large$\frac{1}{2}$}(\ln\lambda)\,x_\alpha\square\Big)
x^\mu
O_\mu(\ka\lambda x, \kb\lambda x)
\big|_{x=\tilde{x}}
\\
\label{O3vec}
O^{\mathrm{tw3}}_{\alpha}
(\ka\xx,\kb\xx)
\!\!&=&\!\!
\int_{0}^{1}\!\!\d\lambda
\Big(\!\delta_\alpha^\mu(x\pd)\!-
\!(x^\mu\pd_\alpha+x_\alpha\pd^\mu)
\!-\!(\ln\lambda)\, x_\alpha\square
x^\mu
\!\Big) O_\mu(\ka\lambda x, \kb\lambda x)
\big|_{x=\tilde{x}}
\\
O^{\mathrm{tw4}}_{\alpha}
(\ka\xx,\kb\xx)
\!\!&=&\!\!
\int_{0}^{1}\!\!\d\lambda\Big(
x_\alpha\pd^\mu+
\hbox{\Large$\frac{1}{2}$}(\ln\lambda)\, x_\alpha\square
x^\mu\Big)
O_\mu(\ka\lambda x, \kb\lambda x)\big|_{x=\tilde{x}}
\end{eqnarray}
\begin{eqnarray}
\label{M_tw2_ir}
M^{\mathrm{tw2}}_{[\alpha\beta]}(\ka\xx,\kb\xx)
\!\!&=&\!\!
2 \int_{0}^{1}\!\!\d\lambda\Big\{\lambda\,
\pd_{[\beta}\delta_{\alpha]}^\mu
-(1-\lambda)\!\big(2x_{[\alpha}\pd_{\beta]}\pd^\mu
-x_{[\alpha}\delta_{\beta]}^\mu\square\big)\Big\}
M_\mu(\ka\lambda x, \kb\lambda x)
\big|_{x=\tilde{x}}
\\
\label{M_tw3_e}
M^{\rm tw3}_{[\alpha\beta]}(\ka\xx,\kb\xx)
\!\!&=&\!\!
\int_{0}^{1}\d\lambda
\Big\{\!\lambda\big((x\pd)\delta_{[\beta}^\nu
- 2x^\nu\pd_{[\beta}\big)\delta_{\alpha]}^\mu
- \hbox{\Large$\frac{1-\lambda^2}{\lambda}$}
\Big(x_{[\alpha}\pd_{\beta]}x^{[\mu}\pd^{\nu]}
\nonumber
\\
&&
+x_{[\alpha}\big(
\delta_{\beta]}^{[\mu} (x\pd)
-
x^{[\mu}\delta_{\beta]}
\big)\pd^{\nu]}
-x_{[\alpha}\delta_{\beta]}^{[\mu}x^{\nu]}\square
\Big)\Big\}
M_{\mu\nu}(\ka\lambda x, \kb\lambda x)\big|_{x=\xx}
\\
\label{M_tw4}
M^{\rm tw4}_{[\alpha\beta]}(\ka\xx,\kb\xx)
\!\!&=&\!\!\!
\int_{0}^{1}\!\!{\d\lambda}
\hbox{\Large$\frac{1-\lambda}{\lambda}$}
\Big\{\!
\big(x_{[\alpha}\delta_{\beta]}^{[\mu}
x^{\nu]}\square
-2x_{[\alpha}\big(
\delta_{\beta]}^{[\mu} (x\pd)
-
x^{[\mu}\delta_{\beta]}
\big)\pd^{\nu]}
\!\Big\}
M_{\mu\nu}(\ka\lambda x, \kb\lambda x)\big|_{x=\xx}
\end{eqnarray}
\begin{eqnarray}
M^{\mathrm{tw2}}_{\alpha}(\ka\xx,\kb\xx)
\!\!&=&\!\!
M_\alpha(\ka\lambda x, \kb\lambda x)
-\xx_\alpha \pd^\mu\!
\int_0^1\!\!\d\lambda\,\lambda\,
M_{\mu}(\ka\lambda x, \kb\lambda x)
\big|_{x=\tilde{x}}
\\
M^{\mathrm{tw3}}_{\alpha}(\ka\xx,\kb\xx)
\!\!&=&\!\!
\xx_\alpha
\Big(\hbox{\Large$\frac{1}{2}$} x^\mu\square
-\pd^\mu(x\pd)\Big)\!\!
\int_0^1\d\lambda\left(\lambda\ln\lambda\right)
M_\mu(\ka\lambda x, \kb\lambda x)\big|_{x=\tilde{x}}
\end{eqnarray}
\hrulefill
\end{table*}

\section{Projection onto the light--cone}

In the preceding Section we stated only the leading
contributions for the different symmetry classes, and
did not carry out the various differentiations
which are necessary to obtain the explicit expressions
for the corresponding operators. Of course, these
differentiations have to be performed before projecting
onto the light--cone. As can be seen easily after that
projection only contributions up to twist $\tau_{\rm max} = 4$
will survive; cf.~Eq.~(\ref{LCdecomp}).
This shows that the nonlocal light--ray operators
under consideration have a unique decomposition which
is given by
\begin{eqnarray*}
\label{O_tw_nl}
O_{\alpha}
\!\!&=&\!\!
 O^{\mathrm{tw2}}_{\alpha}
+O^{\mathrm{tw3}}_{\alpha}
+(O^{\mathrm{tw4}(i)}_{\alpha}
+O^{\mathrm{tw4}(ii)}_{\alpha})
,\\
\label{O_tw_nl}
M_{[\alpha\beta]}
\!\!&=&\!\!
M^{\mathrm{tw2}}_{[\alpha\beta]}
+M^{\mathrm{tw3}(ii)}_{[\alpha\beta]}
+M^{\mathrm{tw4}(ii)}_{[\alpha\beta]}
\\
\!\!&&\!\!
\phantom{M^{\mathrm{tw2}}_{[\alpha\beta]}}
+M^{\mathrm{tw3}(iii)}_{[\alpha\beta]}
+M^{\mathrm{tw4}(iii)}_{[\alpha\beta]}
,\\
M_{\alpha}
\!\!&=&\!\!
\xx^\beta M_{[\alpha\beta]}
=M^{\mathrm{tw2}}_{\alpha}
+M^{\mathrm{tw3}(ii)}_{\alpha}.
\end{eqnarray*}
for the vector and skew tensor cases.
The resulting expressions for general values $(\ka,\kb)$
are put together in the Table below.
To obtain that results some formulas like
$\int_0^1 d\lambda \int_0^1 dt f(\lambda t)/t
= \int_0^1 d\lambda (1-\lambda) f(\lambda)/\lambda$
are used and various partial integrations have to be performed.
The two different twist--4 contributions of the vector operator
may be read off from Eqs.~(\ref{O2vec},\ref{O3vec}) as the
terms multiplied
by $x_\alpha$; analogously for the twist--3 part
$M_{[\alpha\beta]}^{\rm tw3 (ii)}$ of the tensor operator
Eq.~(\ref{M_tw2_ir}).

In conlusion we remark that because of relations (\ref{O2i})
and (\ref{M2ii}) the twist--2 vector and scalar operators and
the skew tensor and vector operators, as well as their anomalous
dimensions, are directly connected. This leads to nontrivial relations
of their non--forward matrix elements, i.e.~the {\em double
distribution amplitudes} and their evolution kernels
\cite{BGR}.
The extension to non--local conformal LC operators is
straightforward.
For a detailed presentation of our results
see Ref.~\cite{GLR}.

\end{document}